\documentclass[twocolumn,letterpaper,10pt]{article}
\usepackage{hicss}
\usepackage{titlesec}
\usepackage{pslatex}
\usepackage{url}
\usepackage{indentfirst}
\usepackage{balance}
\usepackage{parskip}
\usepackage{times}
\usepackage{graphicx}
\usepackage{url} 
\usepackage{subfigure}
\usepackage{paralist}
\usepackage{color}

\titleformat{\section}{\large\bfseries}{\thesection.}{1em}{}
\titleformat{\subsubsection}[runin]{\normalfont\bfseries}{\thesubsubsection.}{3pt}{}

\titlespacing{\section}{0pt}{3ex}{3ex}
\titlespacing{\subsection}{0pt}{2ex}{2ex}
\begin{document}

\title{\vspace*{0.5in}
A Large-scale Analysis of the Marketplace Characteristics in Fiverr}
\author{ \dag Suman Kalyan Maity, \ddag Chandra Bhanu Jha, \ddag Avinash Kumar, \ddag Ayan Sengupta, \ddag Madhur Modi and \dag Animesh Mukherjee \\ \hspace{-2cm}\dag Dept. of CSE, IIT Kharagpur, India - 721302 \hspace{1.5cm} \ddag VGSOM, IIT Kharagpur, India - 721302 \\
Email: \dag\{sumankalyan.maity,animeshm\}@cse.iitkgp.ernet.in; \ddag \{cbj,avinashk2017,ayan.sengupta007,madhurmd\}@iitkgp.ac.in
}

\maketitle
\thispagestyle{empty}

% The abstract is here:
\centerline{{\bf\large Abstract}}
%\vspace{-2ex}

{\it Crowdsourcing platforms have become quite popular due to the increasing demand of human computation-based tasks. Though the crowdsourcing systems are primarily demand-driven like MTurk, supply-driven marketplaces are becoming increasingly popular. Fiverr is a fast growing supply-driven marketplace where the sellers post micro-tasks (gigs) and users purchase them for prices as low as \$5. In this paper, we study the Fiverr platform as a unique marketplace and characterize the sellers, buyers and the interactions among them. We find that sellers are more appeasing in their
interactions and try to woo their buyers into buying their gigs. There are many small tightly-knit communities existing in the seller-seller network who support each other. We also study Fiverr as a seller-driven marketplace in terms of sales, churn rates, competitiveness among various subcategories etc. and observe that while there are certain similarities with common marketplaces there are also many differences.}

\section{Introduction}
\vspace{-5mm}
In recent years, there has been a huge growth in crowdsourcing platforms due to rising demands of crowdsourcing micro-tasks like doing surveys, preparing resumes and other similar tasks that require manual labor. As the popularity of these crowd-based microtasks increased, various crowdsourcing platforms emerged like the Amazon Mechanical Turk, Crowd Flower, Microworker, ShortTask etc. These platforms are primarily demand-driven. Another range of platforms drift from this demand-driven based approach to supply-driven approach where the workers try to showcase their skills and talents by offering various micro-tasks. These marketplaces are primarily used by the freelancers. Upwork, Elance, Fiverr, oDesk, Freelancer, TaskRabbit are some of the very prominent supply-driven marketplaces where the suppliers post advertisements on services like creative writing, programming, software development, graphics etc. with hourly rates or a fixed fee.

Since 2010, Fiverr has grown into one of the largest and popular online freelancing marketplace. As of $11^{th}$ June, 2016, most of the Fiverr traffic (17.9\%) comes from India followed by US (17.6\%) and Pakistan (12.1\%)\footnote{\url{http://www.alexa.com/siteinfo/fiverr.com}}. Fiverr facilitates the buying and selling of ``gigs'' (micro-tasks) online starting from as low as \$5 per gig. The sellers in Fiverr, apart from selling, can also buy gigs from other sellers; similarly, buyers can also become sellers whenever necessary. This dual role played by the members of this community and the interaction among them has made Fiverr a unique marketplace.

\vspace{-2mm}
\section{Related works}
\vspace{-5mm}
There have been several studies on crowdsourcing based online labor marketplaces, predominantly on Amazon Mechanical Turk (AMT). Ross et al.~\cite{ross} presented a user demography study on AMT. Berinsky et al.~\cite{berinsky} described issues related to recruiting subjects on AMT platform. Paolacci et al.~\cite{pao} reviewed the strengths of MT in recruiting participants. Kosinski et al.~\cite{kosinski} measured crowd intelligence. Mason and Suri~\cite{mason} performed behavioral analysis of Turkers and Suri et al.~\cite{honesty} studied their honesty. Halpin and Blanco~\cite{halpin} used machine learning to identify spammers in AMT. Allahbakhsh et al.~\cite{allah} discussed quality control schemes while Heymann and Garcia-Molina~\cite{garcia} proposed a novel analytics tool for crowdsourcing systems. Apart from AMT, Wang et al.~\cite{wangwww} performed analysis of the tasks on ZBJ and SDH (Chinese crowdsourcing sites) and estimated that 90\% of all tasks were crowdturfing tasks.

\textbf{Supply-driven marketplace:} Though there have been several studies on demand-driven marketplaces, supply-driven marketplaces, which are also a growing business in crowdsourcing based marketplaces, are under-studied.~\cite{crowds} performed measurement studies on Fiverr focusing on the quality of the gigs.~\cite{moto} analyzed abusive tasks on Freelancer. Several black-hat marketplaces like HackBB, SilkRoad, Agora, SEOClerks, MyCheapJobs, Gigbucks, Gigton, TenBux have also emerged which facilitate sale of fraudulent products and illicit goods~\cite{christin,pay,serf,lee2014dark}.~\cite{lee2014dark} studied crowdturfing task detection on Fiverr.~\cite{blackhat} analyze marketplace characteristic in SEOClerks and MyCheapJobs.~\cite{liu2014} studied the reward effect on submissions in Taskcn. Our study is different from the above in the following ways - we present a first comprehensive picture of {\em the behavior of the buyers, the sellers and the seller-seller interactions} and demonstrate how all these together impact Fiverr as a marketplace from the {\em economic}, the {\em sociological} and the {\em strategic} perspective.
\vspace{-2mm}
\section{Dataset description}
\vspace{-5mm}
We crawled Fiverr data using preferential crawlers in $R$. \iffalse This involves deploying \textit{RSelenium} package and Chrome browser along with \textit{Chromedriver} as the remote driver facilitator. Clusters are deployed using the Snow package and the task are distributed across workstations.\fi The crawler ran for a period of 20 days and data for 59,786 gigs from 103 subcategories belonging to 10 categories were collected. For each gig, we collected price, reviews, seller's response to reviews, and mutual ratings. We then separately crawled the member profile information (average response time, level of seller, user locations and conversant languages etc.). Next we removed the gigs which have incomplete information for some fields. Consequently, we were left with 41,473 gigs which we use for our study. The dataset contains 21,767 unique sellers and 5,31,841 unique buyers. The total reviews reached a count of 34,43,381 and are used as proxies for business transaction between the reviewer (also buyer) and the gig owner. The prices of the gigs range from \$5 to \$995 indicating the diversity in the service quality offered.\\
\textbf{Gigs, Sellers and Buyers:} Gigs are services provided by sellers in Fiverr and are grouped into categories and further subcategories. 
% `I will fix \textit{Wordpress} Website issues' and `I will advertise your business in my high traffic blog' are typical gigs from the \textit{Wordpress} and \textit{Creative writing} subcategories respectively. Notably, \textit{Wordpress} contributes around 4\% gigs to the market.\\
The \textit{Graphics Design} category has the highest number of sellers with 32\% of the total share. `Level of seller' is assigned by the website to sellers based on the total time spent, the volume of sales, the ratings, the cancellation rates and the order count thresholds amongst many other things. The categories here are \textit{new}, \textit{level 1}, \textit{level 2} and \textit{top rated}. 
% Buyers request gig purchase from sellers. As per our data, around 47\% buyers have purchased gigs from the \textit{Graphic Design} category.
\vspace{-7mm}
\begin{table}[h]
\centering
\caption{Categories based on no. of gigs and buyers.}
\vspace{-2mm}
\label{taboverview}
\resizebox{7.5cm}{!}{
\begin{tabular}{|p{1.5cm}|c|c|c|p{1cm}|p{1cm}|p{1.5cm}|}\hline
 Category Groups & \# sellers & \# buyers & \# gigs & Avg. Price & Avg. Sale & Avg. Revenue\\ \hline
 \textit{Group 1}  & 16962 & 596081 & 20501 & \$31.02  & 93.06 & \$1128.81\\ \hline
 \textit{Group 2}  & 1619  & 67044 & 2304 & \$8.54 & 471.45 & \$3662.11\\ \hline
\textit{Group 3} & 5744 & 87159 & 7394 & \$6.65 & 28.16 & \$218.61\\ \hline
 \textit{Group 4}  & 5012  & 96491 & 6443 & \$5.04  & 35.70 & \$184.80 \\ \hline
 \textit{Group 5} & 3339  & 46670 & 5183  & \$5 & 24.41 & \$122.46  \\ \hline
 \textbf{Total} & \textbf{21767}  & \textbf{531841}  & \textbf{41573} & \textbf{\$8.69}  & \textbf{78.22}  & \textbf{\$757.17} \\ \hline 
\end{tabular}}
\vspace{-4mm}
\end{table}
\\
\textbf{Categories and product differentiation:} We organize all the categories in Fiverr into 5 broad groups -- Graphics \& Design, Digital Marketing, Writing \& Translation; Video \& Animation, Music \& Audio; Programming \& Tech; Advertising , Business; Lifestyle, Gifts, Fun \& Bizarre -- and analyze them separately to find which among them are more popular in terms of the number of gigs, high supply and demand. We also calculate average gig price in each category, average sales of the gigs and revenue generated by the gigs (see Table~\ref{taboverview}). We find that the average price is maximum for the creative categories i.e., group 1 (\$31.02) whereas the average price in the lifestyle related category (group 5) is just \$5. The average sales per gig varies from 471.45 (Audio and Video category) to 24.41 (group 5). We also observe that many categories have significant sales in terms of total sales but show mediocrity in terms of average sales per gig. Therefore, the product differentiation is quite apparent in this marketplace.
\vspace{-2mm}
\section{Fiverr under the economic lens}
\vspace{-3mm}
From the economic viewpoint, we analyze behavior of various marketplace entities like buyers, sellers and the review patterns.
\vspace{-2mm}
\subsection{Behavior of sellers}
\begin{figure}[h]
\begin{center}
\includegraphics[width=0.6\columnwidth,height= 28mm, angle=0]{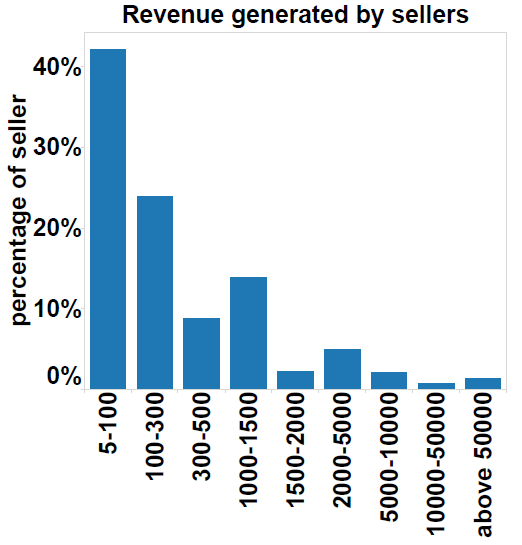}
\vspace{-4mm}
\caption{\label{figrev} Revenue generation of the sellers.}
\vspace{-6mm}
\end{center}
\end{figure}

\begin{figure}[h]
\begin{center}
\includegraphics*[width=0.8\columnwidth,height= 30mm, angle=0]{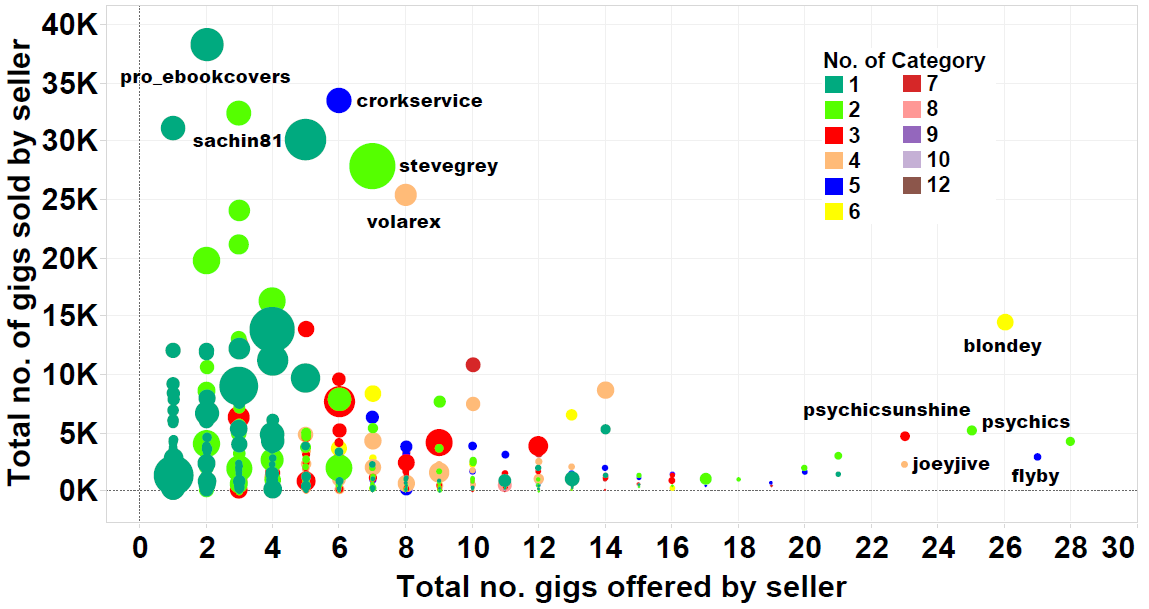}
\vspace{-5mm}
\caption{\label{figtopsale} Sales of top sellers. The sizes of bubbles denotes the amount of total sales. The color coding is based on the number of categories they provide services in.}
\vspace{-7mm}
\end{center}
\end{figure}

\vspace{-2mm}
We find that more than 40\% sellers have only earned \$5-\$100 over the last nine months whereas the proportion of sellers with revenue $>\$50,000$ is only about 0.4\% (see fig~\ref{figrev}).
Fiverr provides a lot of information about sellers and gigs e.g., average response time, orders in queue, number of reviews, number of positive reviews, average response time of the seller and many recently added new features like - a rating based on the service as described on a scale of 0--5, a rating based on the recommendation of the gig by a user. We investigate several of these features in order to unfold the behavior of the sellers. 
% Also, many sellers sell more than one gig even in the same subcategory. The intent may be to maximize their visibility. Figure \ref{fig3} shows the histogram plot for number of gigs provided by sellers.
% \begin{figure}[h]
% \begin{center}
% \includegraphics*[width=1\columnwidth,height= 45mm, angle=0]{seller_descripiton.PNG}
% \caption{\label{fig3} The above two plots shows histogram plot for number of gigs provided by sellers and number of categories where they sell based on their level. The below two show the same study done for the overall population.}
% \end{center}
% \end{figure}
Based on different attributes, Fiverr classifies the sellers into three levels - \textit{level 1}, \textit{level 2} and \textit{top rated}. There are many sellers who have not yet met the criterion to be in any of these levels and are called \textit{new} sellers by Fiverr. We observe that about 22.56\% of the whole seller community are in \textit{level 1}, more than 62.89\% are \textit{level 2} sellers whereas only 4.8\% are \textit{top rated} sellers and the rest of them are \textit{new}. Table~\ref{tabsalerev} shows how sales and revenue varies for the different levels of sellers. Apart from the fact that the \textit{top rated} sellers are selling more number of products, it is also quite clear that they charge more for their product than others. A \textit{top rated} seller charges \$9.56 on average, where a \textit{level 2} seller charges \$8.96. This premium charging shows their bargaining power over others.
\vspace{-4mm}
\begin{table}[h]
\centering
\caption{Performance of sellers.}
\label{tabsalerev}
\vspace{-3mm}
\resizebox{7cm}{!}{
\begin{tabular}{llll}
 \textbf{Level}& \textbf{Avg. Sale} & \textbf{Avg. Revenue}& \textbf{Avg. Gigs}\\
\textit{Level 1} & 12.94 & \$112.66 & 1.46 \\
\textit{Level 2} & 158.22 & \$1417.15 & 2.22 \\
\textit{Top rated} & 2344.24 & \$22,413.26 & 4.04    
\end{tabular}}
\vspace{-4mm}
\end{table}
\\
\textbf{Sales and top sellers }
First we observe the performance of sellers based on the revenue earned in the last nine months. The revenue generated by the seller varies from \$555,715 to only \$5. If we consider the number of unit sales without considering the price, then it can be as high as 38,302 to as low as 1. From the fig~\ref{figtopsale}, we can clearly observe that most of the top sellers prefer selling a small number of gigs in few subcategories rather than providing diverse type of gigs. Hence, they are mostly proficient in certain type of services which indicates a good level of professionalism in a service driven marketplace.
% \begin{table}[h]
% \centering
% \caption{Top sellers: based on revenue.}
% \label{tabrev}
% \vspace{-2mm}
% \resizebox{7cm}{!}{
% \begin{tabular}{lllll}
% \textbf{Username} & \textbf{Level} & \textbf{Sales} & \textbf{Revenue} & \textbf{No. of Gigs} \\
% stevegrey & Top rated & 27,887 & \$5,57,155 & 7 \\
% Woofy31 & Top rated & 13,866 & \$5,37,020 & 4\\
% sachin81 & Top rated & 30,103 & \$4,48,635 & 5\\
% alinam1 & Top rated & 1,406 & \$4,41,770 & 1\\
% volunteer & Top rated & 9,057 & \$3,92,470 & 3\\
% \end{tabular}}
% \end{table}

% \begin{table}[h]
% \centering
% \caption{Top seller: based on sales figure.}
% \label{tabsale}
% \vspace{-2mm}
% \resizebox{7cm}{!}{
% \begin{tabular}{lllll}
% \textbf{Username} & \textbf{Level} & \textbf{Sales} & \textbf{Revenue} & \textbf{No. of Gigs} \\
% pro ebookcovers & Top rated & 38,302 & \$2,87,265 & 2 \\
% crorkservice & Top rated & 33,503 & \$1,67,515 & 6\\
% blboss & Level 2 & 32,374 & \$1,61,870 & 3\\
% amazesolutions & Top rated & 31,135 & \$1,55,675 & 1\\
% sachin81 & Top rated & 30,103 & \$4,48,635 & 5\\
% \end{tabular}}
% \vspace{-5mm}
% \end{table}

\begin{figure}[h]
\begin{center}
\includegraphics*[width=0.8\columnwidth,height= 28mm, angle=0]{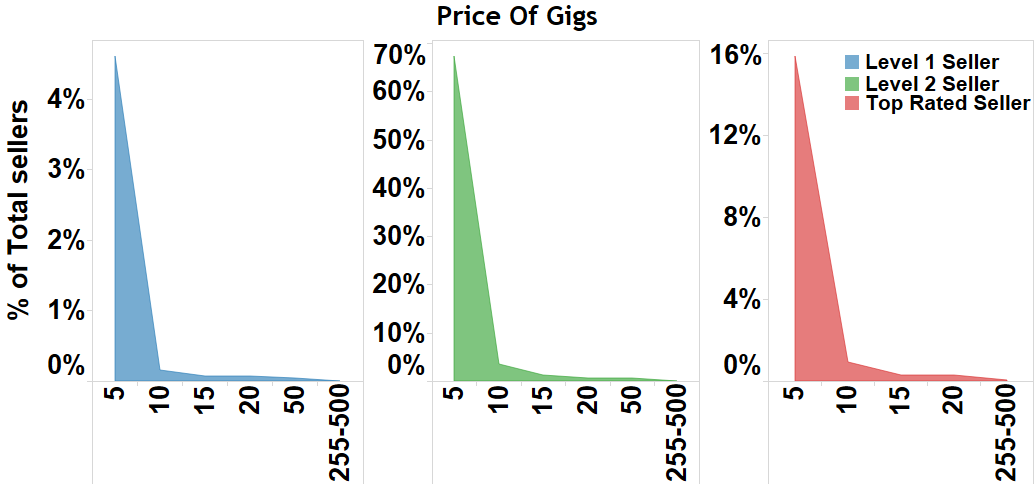}
\vspace{-2mm}
\caption{\label{figprice} Price of gigs of different sellers in each level.}
\vspace{-6mm}
\end{center}
\end{figure}

\textbf{Pricing of gigs }
In Fiverr, 87.73\% of the gigs are priced at \$5. 0.02\% of the gigs are highly priced and more than \$500. From the dataset, we observe that most of the highly priced gigs are from the subcategories \textit{Buy Photos Online Photoshop}, \textit{Banner Ads} and \textit{Content Marketing}. Figure~\ref{figprice} shows how different levels of sellers price their gigs. The underlying distribution within each seller level appears to be the same with the only difference being the percentage of total sellers in level 2 is much high compared to other levels.

\begin{figure}[h]
\begin{center}
\subfigure[]{\label{figrate}\includegraphics[width=0.48\columnwidth, angle=0]{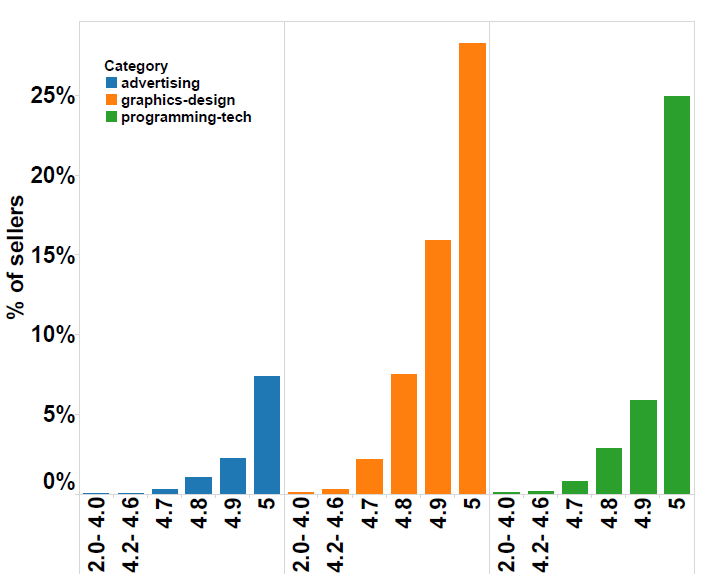}}
\subfigure[]{\label{figresponse}\includegraphics[width=0.48\columnwidth,height= 35mm , angle=0]{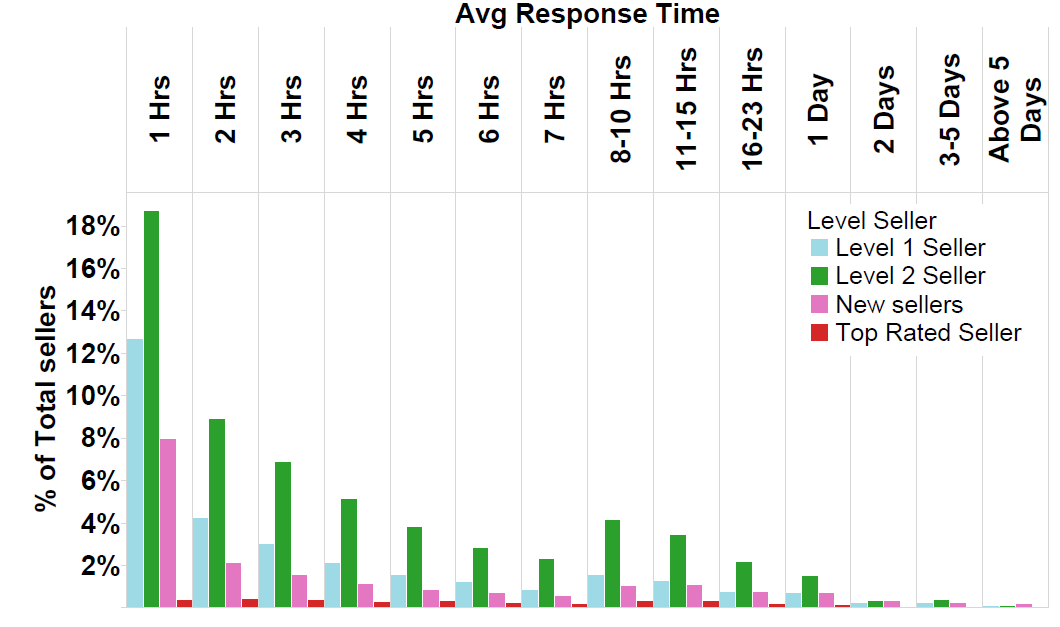}}
\end{center}
\vspace{-4.8mm}
\caption{(a) Average buyer rating of sellers for different categories (b) Average response time of seller.}
\label{fig:rate_response}
\vspace{-4mm}
\end{figure}

\begin{figure}[h]
    \centering
\begin{minipage}[b]{0.42\columnwidth}
\includegraphics[width=1\columnwidth,height= 30mm, angle=0]{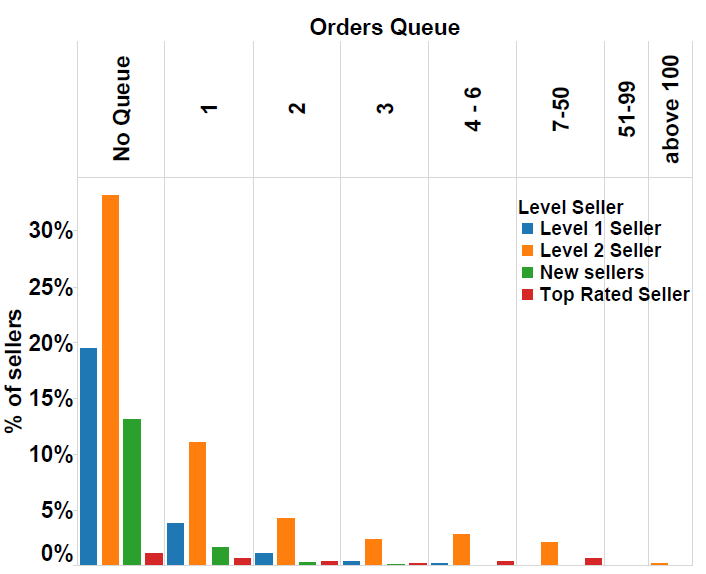}
\caption{\label{figorder} Distribution of order queue count of sellers.}
\end{minipage}
\begin{minipage}[b]{0.42\columnwidth}
\includegraphics[width=1\linewidth,height= 30mm, angle=0]{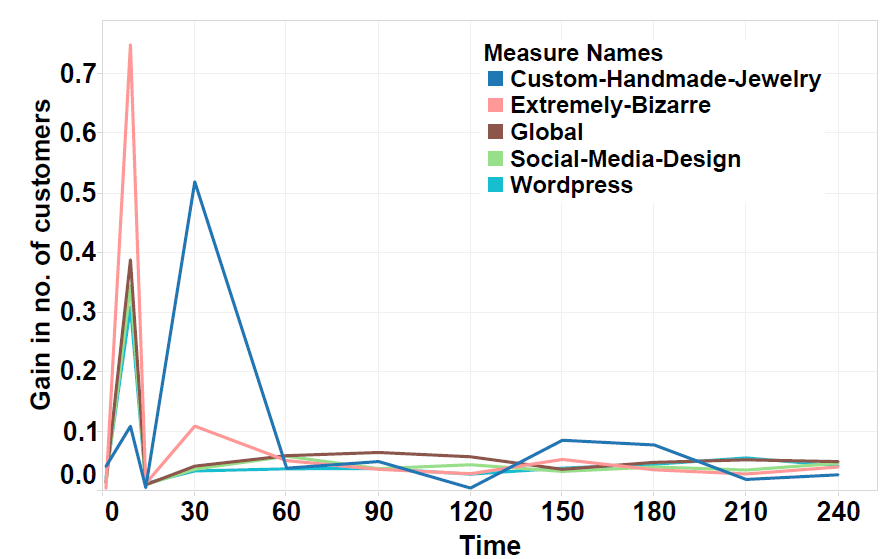}
\caption{\label{figgain} Gain of new customers by sellers in each subcategory.}
\end{minipage}
\vspace{-6mm}
\end{figure}

\textbf{Buyer ratings and seller response times }
Buyers provide ratings to the sellers for each purchase on a scale of 0-5. Hence, average rating for each seller can be a parameter for judging his/her performance. After analyzing the data, we believe that the rating system is not very discriminatory for Fiverr. More than 60\% of the sellers have an average rating of 5 and more than 96\% have more than 4.7. This distribution is also valid category wise (see fig~\ref{figrate}). %\todo{Why not show this for the different categories of sellers?} 
In fig~\ref{figresponse}, we show the distribution of average response time of sellers. The graph shows power-law behavior. More than 95\% of the sellers usually respond within a day and $\sim35\%$ of them respond within an hour.
\\ \textbf{Order queue }
Another new feature that Fiverr has recently added is the order queue count which tells us how many incomplete orders are present for a particular seller for a given gig. Low value of order queue count can be either because of less popularity or excellent customer handling. Similarly, high value of order queue count can indicate less proficiency in customer handling or large number of orders inflow. In fig~\ref{figorder}, we show the distribution of no. of orders in queue for each level of sellers. Most of the \textit{level 2} sellers are very prompt, however, a non-negligible percentage of them also delay their services. 
% \subsection{Peak time}
% Next we studied the peak time in last 9 months for each of the sellers to see the whether peak time is random or, there is some particular pattern among their peak time. For each seller we define
% \begin{center}
% $peak$ $time$ = $S_j$ with $j$ = $argmax_i$ $sales_i/T_i$
% \end{center}
% Where, $sales_i$ is total sales of the seller at $i$th time period, $T_i$ is the length of the time period in days and $S_i$ is starting day number of $i$th time period. Hence, it is essentially the time frame for which the average sale per day is maximum. For most of the subcategories, we observe in figure \ref{figpeak} that the peak time is around the last week of the current month i.e. around the $30$th day. This can be justified by the fact that most of the popular or, front paged sellers have done good in the current month period. Because of that, Fiverr is promoting their gigs in the front page of its website.
% \begin{figure}[h]
% \begin{center}
% \includegraphics*[width=0.85\columnwidth,height=35mm, angle=0]{peak_time.PNG}
% \caption{\label{figpeak} Line plot of seller's peak time in days.}
% \end{center}
% \end{figure}

\textbf{Gain of new customers}
The customer base of sellers keeps changing over time and the inflow of new customers can be taken as a proxy for growing popularity and hence an indicator of good services. We propose a metric to measure this inflow. For a particular seller, we define
\begin{center}
$Gain(t) = \frac {|customer_{t} \cap  \overline{customer_{t-1}}|} {|customer_{t-1}|}$
\end{center}
where, $customer_{t}$ is the set of all customers of the seller at time $t$ and $\overline{customer_{t}}$ is the set complement of the set $customer_t$. $customer_{t-1}$ represents the set of customers one day before the $t^{th}$ day. 
Fig~\ref{figgain} shows the average gain of top sellers of five subcategories. We observe that average gain decreases over time. Globally, the percentage gain of customers for top sellers is around 40\% and is greater than the preceding time intervals for one month. Some subcategories show different trends. For example - in \textit{Custom Handmade Jewelry} the largest gain ($> 50$\%) of customers took place during the first week of May 2016. Similarly, around the middle of May, 2016 the subcategory \textit{Extremely Bizarre} showed the maximum gain of new customers which is around 75\% (fig~\ref{figgain}).
\vspace{-2mm}
\subsection{Behavior of buyers}
\begin{figure}[h]
\begin{center}
\subfigure[]{\label{figbuygigs}\includegraphics[width=0.48\columnwidth, height= 25mm, angle=0]{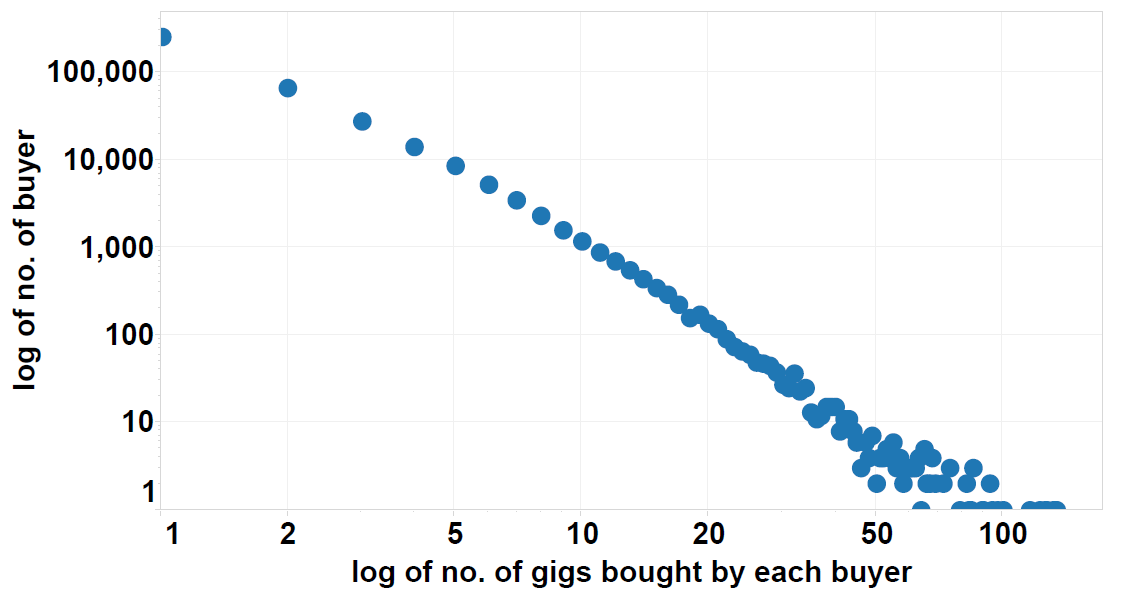}}
\subfigure[]{\label{figbuysub}\includegraphics[width=0.48\columnwidth, height= 25mm, angle=0]{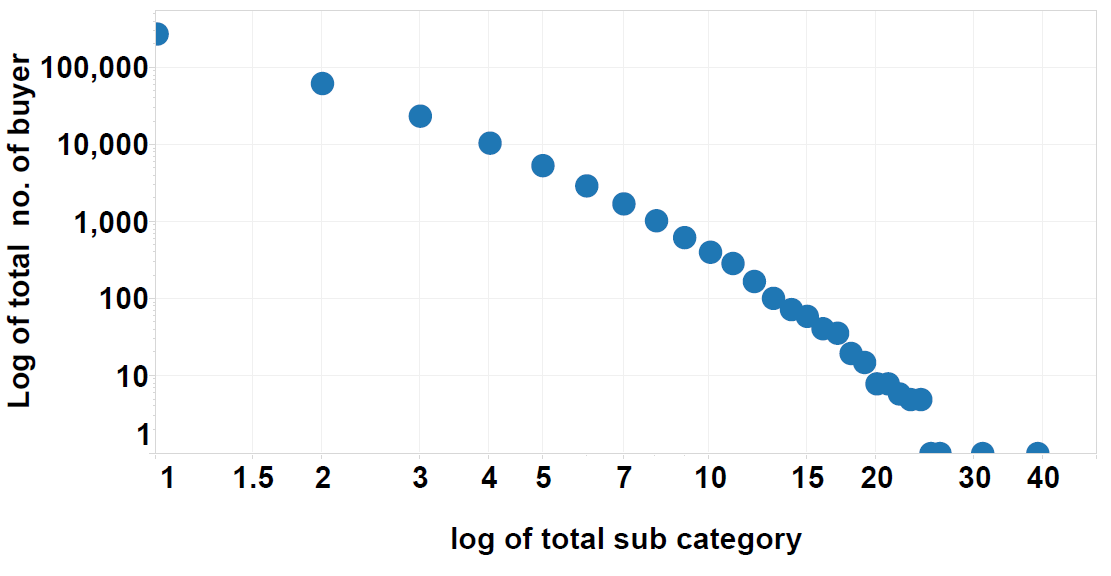}}
\vspace{-3mm}
\caption{\label{figrel} Distribution of (a) no. of gigs (b) no. of subcategories from which the buyers buy gigs.}
\vspace{-5mm}
\end{center}
\end{figure}

\begin{figure}[h]
\begin{center}
\includegraphics*[width=0.7\columnwidth,height= 25mm, angle=0]{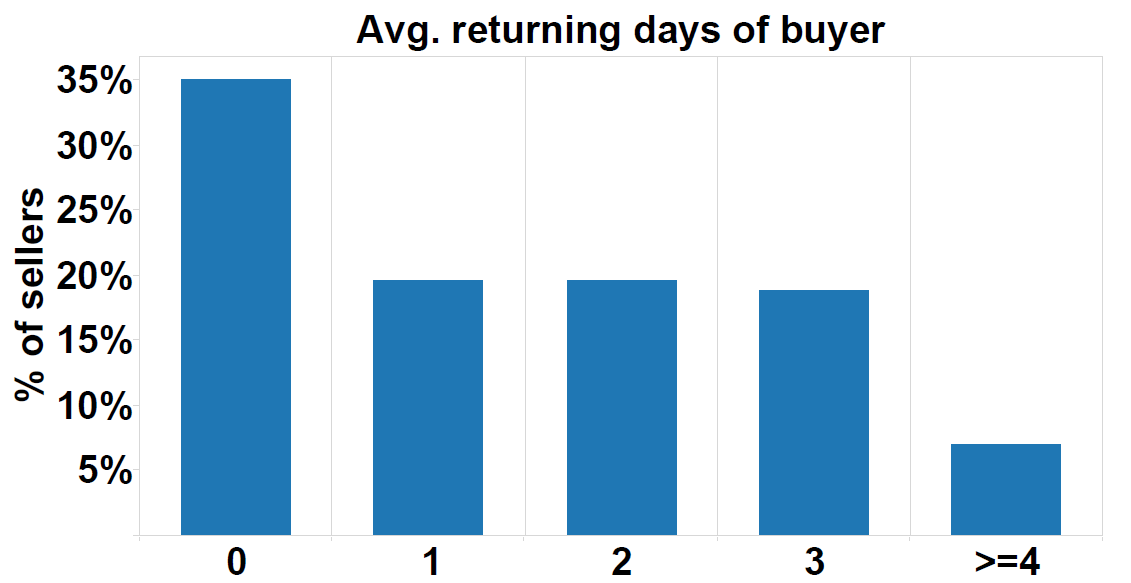}
\vspace{-2mm}
\caption{\label{figreturn} Distribution of returning time of buyers.}
\vspace{-7mm}
\end{center}
\end{figure}

\vspace{-3mm}
Buyers are the driving force of any marketplace. In our dataset, we have a total of 5,31,841 buyers. Nevertheless, if we only want to observe the profitable or returning customers, then the number would be much less than this. $\sim45\%$ of the buyers are one-time (only one purchase across all gigs) buyers. Only 18\% of the buyers have purchased more than 5 times in last nine months, with the maximum being 2939. Fig~\ref{figbuygigs} shows the distribution of number of purchases of gigs indicating a power-law behavior. In fig~\ref{figbuysub}, we present the distribution of purchases from different subcategories. 68.62\% of the buyers buy from one subcategory and only nine buyers have bought from more than 30 subcategories.
\vspace{-4mm}
\paragraph{Purchase and top buyers}
As we have seen that most of the buyers are just one time buyers, we consider only the top buyers in our studies, whose total purchases are more than 200.
% We present the statistics of some of the top buyers in table~\ref{tabbuyp} and~\ref{tabbuyn}. We find that some of the observations are very striking in terms of common service-based marketplaces. For example - four out of the five buyers have bought just one gig in last nine months multiple times. In all the cases the number of purchases is more than 525, which amounts to more than 1.944 purchases per day on average, which is very unusual.
% \begin{table}[h]
% \centering
% \caption{Top buyers: based on purchase amount.}
% \label{tabbuyp}
% \vspace{-2mm}
% \resizebox{7.5cm}{!}{
% \begin{tabular}{lllll}
% \textbf{Username} & \textbf{Gigs} & \textbf{Revenue} & \textbf{Avg. purchase} & \textbf{Categories} \\
% \textit{sonokraft} & 23 & \$41,460 & 19.52 & 11 \\
% \textit{sherlockholmes} & 7 & \$31,515 & 9 & 4\\
% \textit{avamallory} & 5 & \$30,025 & 13 & 3\\
% \textit{fictionbuyer} & 5 & \$21,080 & 11.6 & 3\\
% \textit{revansylad} & 10 & \$20,115 & 6.3 & 7\\
% \end{tabular}}
% \vspace{-4mm}
% \end{table}
% 
% \begin{table}[h]
% \centering
% \caption{Top buyers: based on average number of purchases.}
% \label{tabbuyn}
% \vspace{-2mm}
% \resizebox{7.5cm}{!}{
% \begin{tabular}{lllll}
% \textbf{Username} & \textbf{Gigs} & \textbf{Revenue} & \textbf{Avg. purchase} & \textbf{Categories} \\
% \textit{rimizabe} & 4 & \$14,690 & 734.5 & 4 \\
% \textit{jeffshaw291} & 1 & \$3,315 & 663 & 1\\
% \textit{georginacutler} & 1 & \$4,750 & 570 & 1\\
% \textit{verykcniearkmi8} & 1 & \$2,850 & 570 & 1\\
% \textit{tara555} & 1 & \$4,375 & 525 & 1\\
% \end{tabular}}
% \vspace{-4mm}
% \end{table}
\begin{figure}[h]
\begin{center}
\includegraphics*[width=0.8\columnwidth,height= 33mm, angle=0]{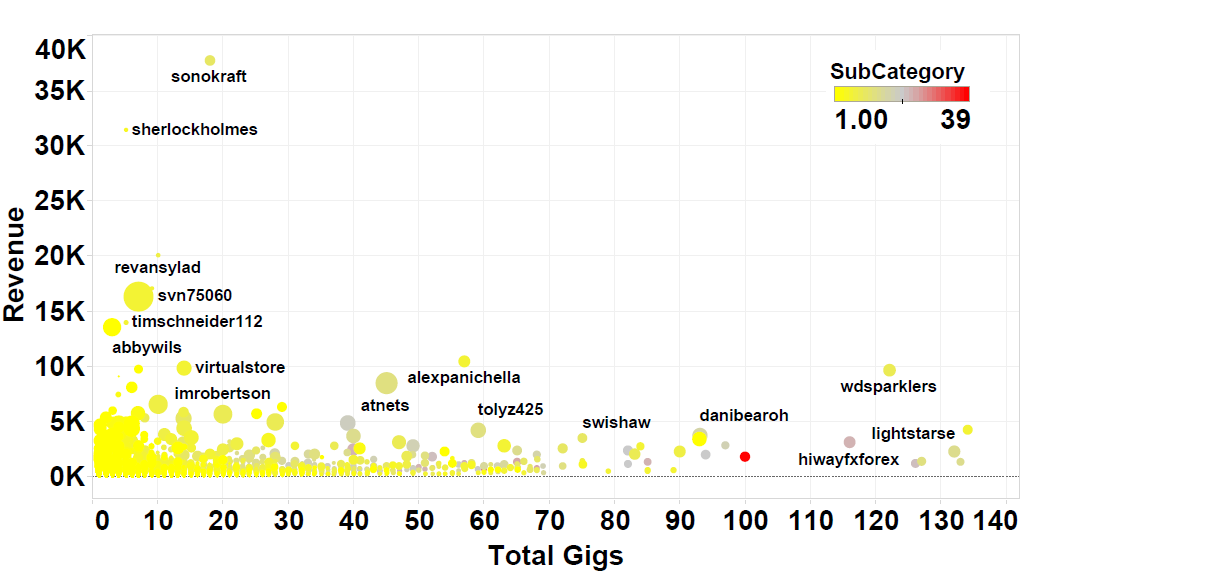}
\vspace{-5mm}
\caption{\label{figpurchase} Purchase plot of buyers. Size indicates the number of gigs a buyer has bought. Color coding is based on the number of subcategories the buyer is buying from.}
\vspace{-7mm}
\end{center}
\end{figure}
In fig~\ref{figpurchase} we show that the amount of purchases is not always directly proportionate to the volume of purchases. Moreover, most of the top buyers prefer to buy small number of gigs from a few subcategories, rather than buying different types of products across different subcategories. This also shows that these buyers are not just stray visitors to the website; instead they repeatedly buy the same products over time. Hence, it is legitimate to assume that loyal customers (although less in number) do exist in supply-driven marketplaces like Fiverr. 
\vspace{-2mm}
\paragraph{Returning time of buyers}
As we have seen the existence of loyal and repeated buyers in Fiverr, the next question that arises is - how often does a returning customer buy a particular gig? To measure this, we define a statistic for each of the repeated buyers for each of the gigs they are buying. 
\begin{center}
$returning$ $time$ = $\sum_{i=2}^{N}{\frac{|T_i - T_{i-1}|}{N-1}}$
\end{center}
where, $\{T_i\}_{i=1}^N$ is the time periods of purchase of a particular gig. In other words, this statistic measures the average day difference between any two consecutive purchases for each of the repeated buyers. In fig~\ref{figreturn} we show the distribution of the \textit{returning time} of the top buyers in Fiverr. It clearly shows the peak at time zero and almost uniform frequencies up to 3rd day followed by a decrease. We can therefore infer that most of the frequent buyers usually return \textit{within a day}, which is another unlikely event for other e-commerce and product driven marketplaces. This observation can greatly help the sellers to incentivise returning customers. This also shows the extent of heavy usage of services by the frequent customers which is overall an encouraging sign for Fiverr. 
\color{black}
\vspace{-5mm}
\subsection{Review characteristics}
\begin{figure}[h]
\begin{center}
\subfigure[]{\label{figrevuser} \includegraphics[height=30mm,width=0.45\columnwidth, angle=0]{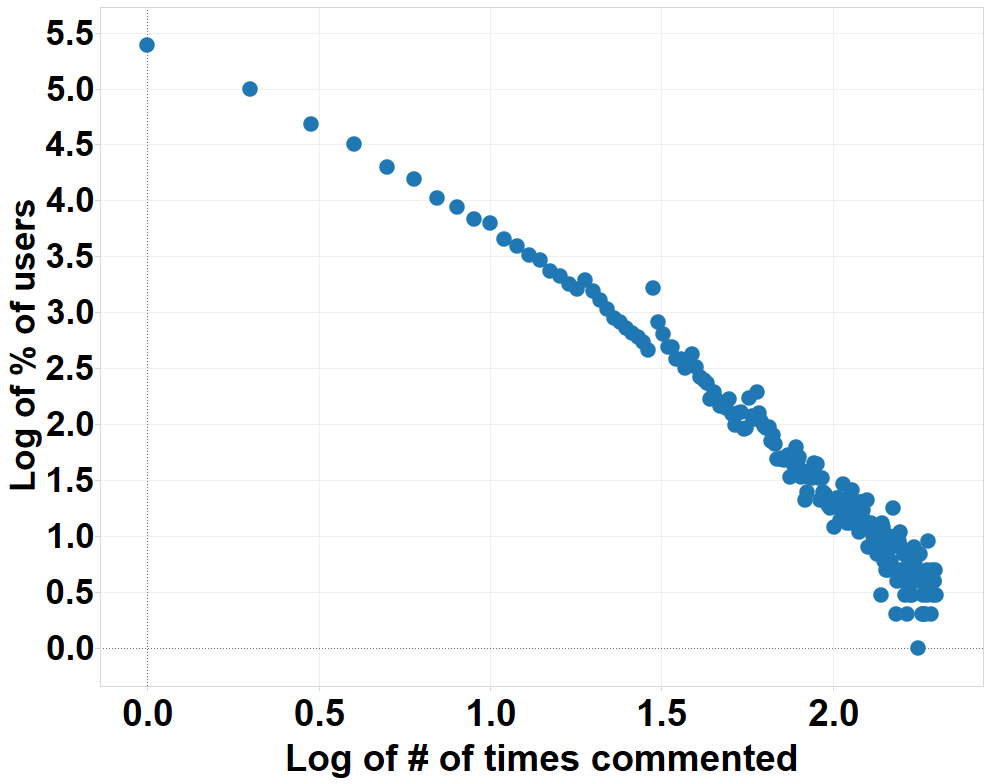}}
\subfigure[]{\label{figrevgig} \includegraphics[height=27mm,width=0.45\columnwidth, angle=0]{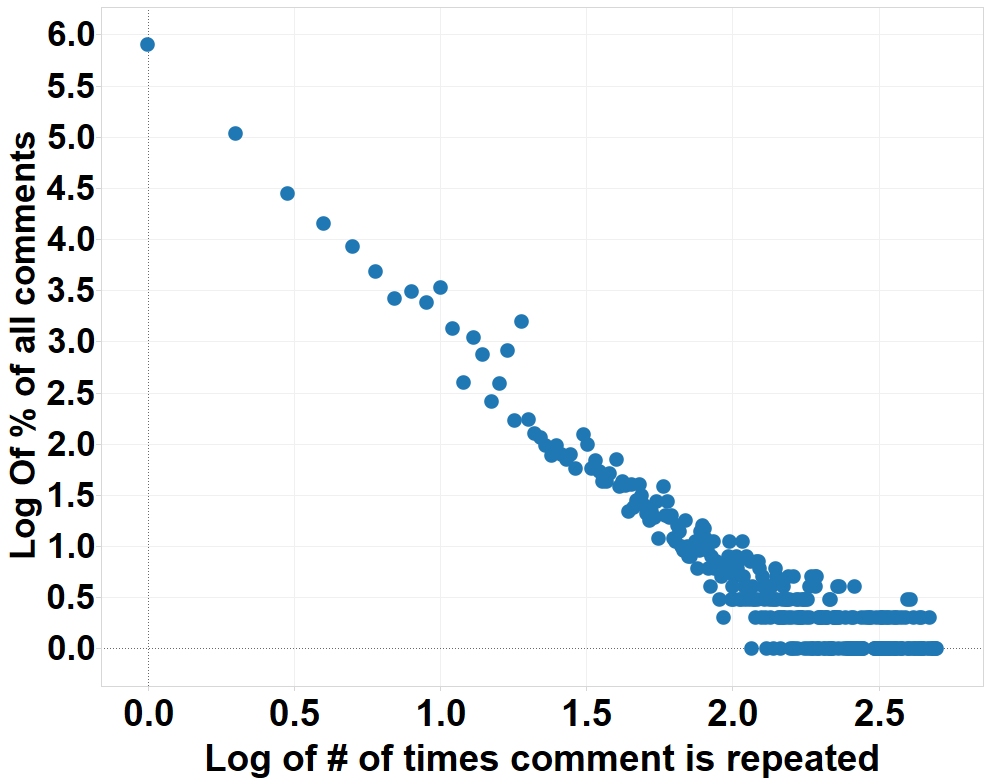}}
\vspace{-3mm}
\caption{\label{figreview} Distribution of (a) number of times the same comment is used by different users (b) number of times the same comment is used in different gigs.}
\vspace{-7mm}
\end{center}
\end{figure}

\vspace{-2mm}
Reviews have increasingly become one of the guiding foundations for buyers in online marketplaces. \iffalse With the constant bombardment of new products and features everyday on e-commerce sites, making the right choice often boils down to taking clues from the existing reviews. In such a scenario, it becomes pertinent that\fi A good review system allows for a discriminative analysis among the different choices and helps the consumer determine the usefulness and ingenuity of the product under question in the different stages of the purchase decision making process.
\iffalse Amazon has a well developed review system~\cite{mudambi} and customers use this regularly to compare products. Relevant reviews are promoted based on customer rating, review timing, content, ingenuity and other latent factors. The idea is to retain the ones which contain relevant information.\fi Fiverr's review system, though probably intended to be means for insights into gig's performance, does seem to be quite different from the usual e-commerce platforms. The buyer and the seller can engage in a mutual review exchange and rate each other based on their interaction. Initial analysis shows that the linguistic structure has evolved along a very constrained framework which involves a high degree of mutual appreciation, well-wishing and goodwill exchange.
We perform general analysis of gig reviews and try to look at the commenting behavior of the users across all the gigs. Though we consider comments on a gig as a proxy for purchase, we study the repetitiveness of comments in Fiverr. The normalized figures for 5,31,841 users show us that $\sim45\%$ of the users comment only once. This is followed by 18\% of the users placing the same comment twice. $\sim8.9\%$ of the users have commented the same thing thrice. This distribution follows a power-law behavior (see fig~\ref{figrevuser}).
We also observe how comments are being repeated across all the gigs. Figure~\ref{figrevgig} shows the distribution of the comment repetitions which shows a power-law curve with heavy tail. 80\% of the comments are non-repetitive, 10.98\% have occurred twice. Notably, the single comment ``Outstanding Experience!'' has appeared 51,451 times across all comments. Close variations of this comment are also observed frequently. This is followed by 12,406 repetitions of the comment ``thanks''. \\ 
\textbf{Sentiment analysis:} We perform sentiment analysis of the reviews using the standard sentiment dictionary~\cite{senti}. Most of the reviews have positive and encouraging constructs. This can be attributed to the structure which has developed over time and is peculiar to Fiverr. ``Outstanding experience'', ``Thanks'' and ``Great to work with'' are some of the most common phrases encountered and this is reflected in the skewed sentiment scores. However, a difference in the polarity distribution can be observed across categories with different inherent characteristics. The difference is more pronounced across different price ranges.
\begin{figure}[h]
\begin{center}
\subfigure[]{\label{figsentibuy}\includegraphics[height=30mm,width=0.45\columnwidth, angle=0]{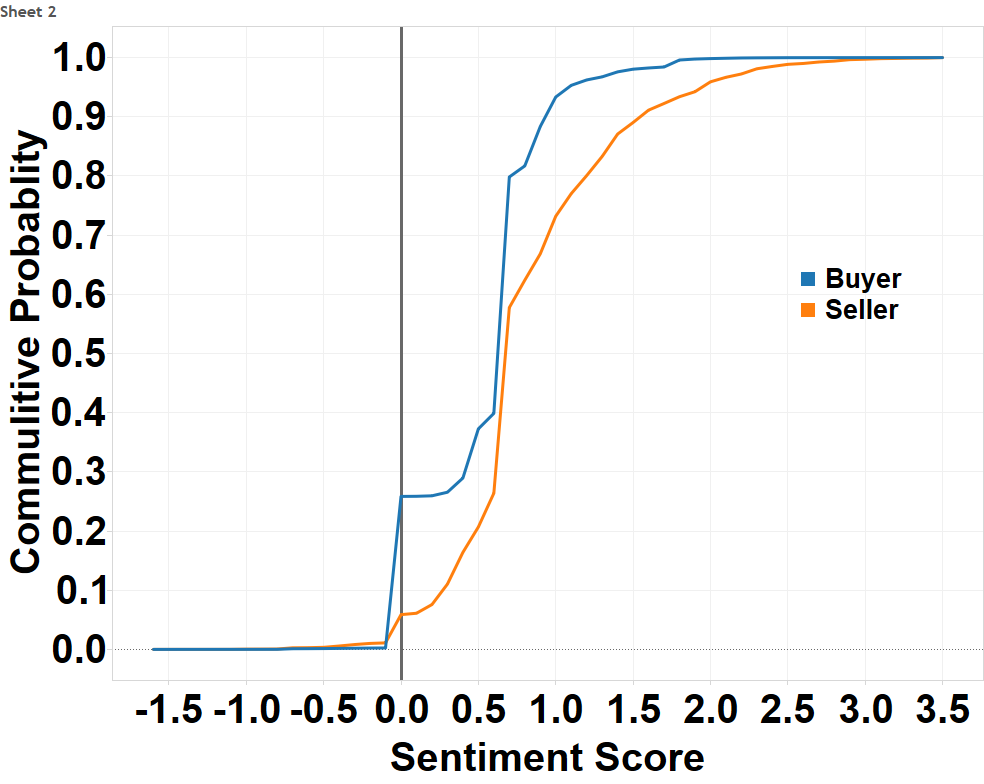}}
\subfigure[]{\label{figsentisub}\includegraphics[height=30mm,width=0.45\columnwidth, angle=0]{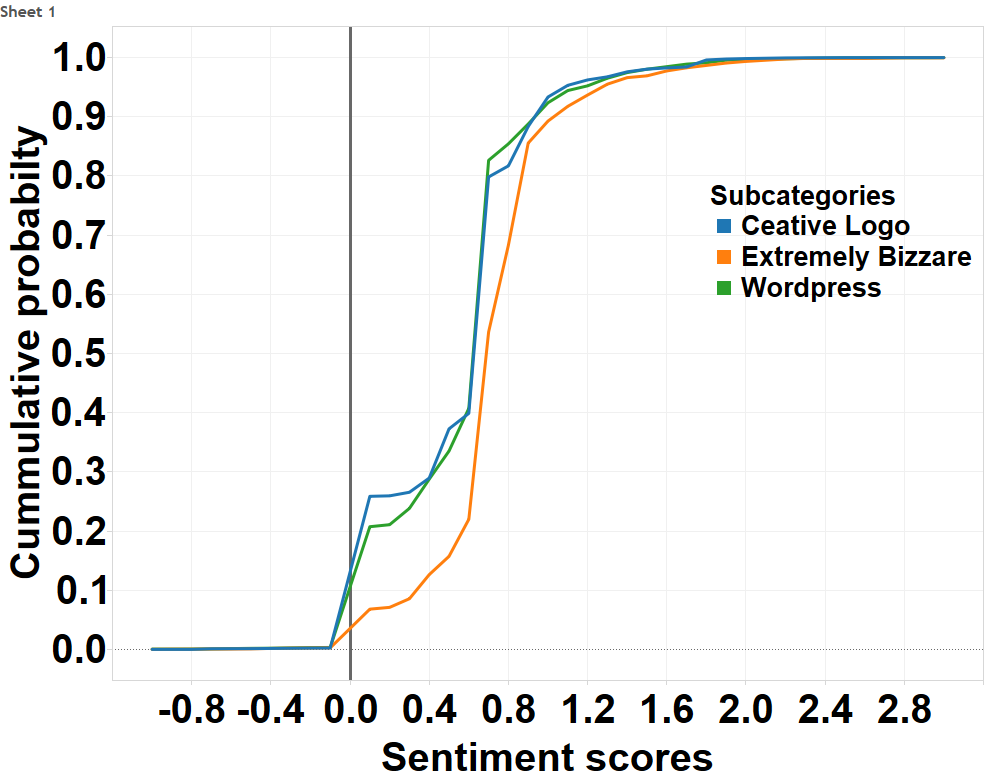}}
\vspace{-5mm}
\caption{\label{figsenti} (a) Comparison of sentiment of buyer and seller text exchange for the Creative Logo Design subcategory (b) Review sentiment comparison across three categories.}
\end{center}
\vspace{-5mm}
\end{figure}
Sellers are found to be more positive in their responses to the comments by the buyers. A cumulative plot of Creative Logo Design subcategory demonstrates this fact (see fig~\ref{figsentibuy}). Comparisons across different subcategories (see fig~\ref{figsentisub}) also gives very convincing insights. For trivial services like the ones contained in Extremely Bizarre, the reviews have higher proportion of positive sentiments than more serious services like the ones in \textit{Wordpress} and \textit{Creative Logo Design}.\\
%\subsubsection*{Price sensitivity}
%The relation between the buyer’s feedback and the corresponding seller response is of special interest. A close examination of the plots over different categories brings out the contrast. Categories portraying relative static skills of sellers, like \textit{Cartoons and Caricatures} and those where the underlying skill is actively evolving, e.g. - \textit{Programming and Tech} show particularly distinct behavior in this interaction. The former shows more similarity in the exchange as compared to the latter where the gigs are more likely to be costlier and non-personal in nature. \textit{Photoshop} subcategory has gigs priced as high as \$995 and expression of dissatisfaction in service is more in this case as compared to \$5.
\textbf{Comparison with Amazon's reviews:} Since reviewing patterns are a key representative of any marketplace, we use these to differentiate a supply-driven and a demand-driven marketplaces. We collected Amazon's review data for the \textit{Electronics} category and randomly sampled the reviews. These are in high contrast to that of Fiverr. The sentiment distribution is more inclusive of negative polarities and the linguistic structure looks more realistic upon primary inspection. Comparison of top 10 most frequent words (see table~\ref{tabletop10}) in reviews of Amazon \textit{Electronics} products and Fiverr's most sold \textit{Creative Logo Design} shows that Amazon's products get feedback related to product specifics with some words expressing gratefulness while Fiverr is dominated by mostly emotional responses. The average number of characters per review for Amazon is around 381, while for Fiverr, it is merely around 61. This can be attributed to the elaborate and explanatory nature of the former and a generic congratulatory nature of the later.
\vspace{-4mm}
\begin{table}[h]
\centering
\caption{Top 10 most frequent words in reviews}
\vspace{-4mm}
\label{tabletop10}
\resizebox{7.5cm}{!}{
\begin{tabular}{ll}
 \textbf{Marketplace} & \textbf{Top 10 words}\\
 Amazon  & case phone one great works like just well good use\\ 
 Fiverr & experience great work outstanding thanks logo job \\
  & will good excellent
\end{tabular}}
\vspace{-4mm}
\end{table}
\vspace{-3mm}
\section{Sociological/network perspective of Fiverr}
\vspace{-3mm}
\iffalse In this section, we study the interactions among the sellers in terms of sales and purchase to gain insights about their behavior as a community. Towards this objective,\fi We create two types of seller-seller networks (Tier-I and Tier-II) and study the interesting properties of these networks.
\vspace{-2mm}
\subsection{Tier-I seller-seller network}
\begin{figure}[h]
\begin{center}
\includegraphics*[scale=0.2, angle=0]{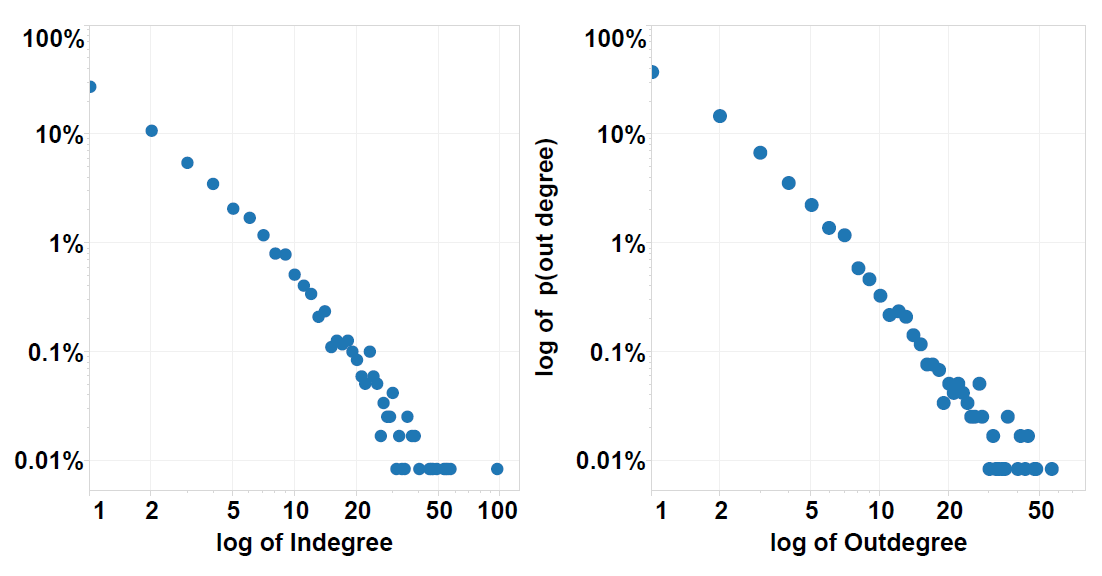}
\vspace{-4mm}
\caption{\label{figdeg} Degree distribution of the network.}
\vspace{-4mm}
\end{center}
\end{figure}

\vspace{-2mm}
%\todo{Since there are two types of networks constructed I would recommend that you name them differently. Otherwise it is too confusing. Call the first one ``First tier/level seller-seller network'' and the second one as ``Second level/tier network.''}
We create a Tier-I seller-seller network $G = (V,E)$ as follows. The set of nodes $V = \{s_1, s_2, ...\}$ are the set of sellers in Fiverr and if a seller $s_1$ buys $k$ times from other seller $s_2$, then we add a directed edge from $s_2$ to $s_1$ in $G$ with weight (proportional to no. of purchases)
\begin{center}
$weight(s_2, s_1)$ = $k/total\_sales(s_2)$
\end{center}
\textbf{Graph density:} The network formed using the above method has 49.13\% of the full sellers community but with a graph density of only 0.001. This low number indicates that, although a number of sellers together form the network, they are loosely connected to each other. Thus they usually have a very small set of sellers in their neighborhood from whom they make a purchase.\\
\textbf{Degree distribution and power law properties:} It is a known fact that degree distribution in most of the real world networks follows a power law \(p(k) \approx k^{-\alpha} \) with the exponent $\alpha$ lying between 2 and 3. The Fiverr seller-seller network also exhibits the same characteristics with $\alpha$ value of 2.27 for in-degree and 2.41 for out-degree (see fig~\ref{figdeg}). We further observe that percentage of nodes with zero in-degree is 43.53\% whereas percentage of nodes with zero out-degree is 29.45\%. This shows that the number of sellers who are only buying from other sellers is much higher compared to those who are only selling.\\
\textbf{Transitivity:} Transitivity, also referred to as the clustering coefficient, is the measure of how well the neighbors of a node are connected among themselves. This also indicates the existence of cliques or dense modules within a network. The Tier-I seller-seller network of Fiverr has very low clustering coefficient of 0.037 (the direction of the edges are ignored in this computation). The low clustering coefficient shows that there is no tightly connected module existing between the sellers in the Fiverr market which is in contrast to the general market where it is usually known to exist. \\
\textbf{Connected components and communities:} The network consists of 526 weakly connected components and 11,077 strongly connected components. The number of strongly connected components is almost same as the number of nodes which show low reciprocity in the network. By reciprocity of a network $G(V,E)$, we understand the proportion of occurrences where both $(v_1,v_2)$ and $(v_2,v_1)$ belongs to the edge set $E$. Using Louvain algorithm~\cite{comm}, we determine the community structure of this network. 625 communities are discovered with an overall high modularity value of 0.941. Therefore, the network shows many small communities and the communities themselves are tightly-knit.
%\begin{figure}[h]
%\begin{center}
%\includegraphics*[scale=0.1, angle=0]{modularity.png}
%\caption{\label{figcomm} Top five communities based on the number of members.}
%\vspace{-4mm}
%\end{center}
%\end{figure}
\\
\textbf{Geography of sellers:} We collected the geographical location based data of the sellers associated with the Tier-I seller-seller network. We find that 31.57\% of the whole community belongs to United States, followed by Pakistan (9.27\%) and India (7.73\%). Moreover, the order remains same for individual sellers and buyers in the community, where United States is leading followed by India and Pakistan. We also observe that in 26.15\% of the cases both the buyer and the seller belong to the same country.\\
\textbf{Characterizing the Tier-I seller-seller network:} We observe that in 15.53\% cases a seller buys from another seller who sells gigs in the same category itself. Now, if a buyer, who himself sells gigs in the same subcategory buys something similar to his product, then the explanations behind this could be either promoting the other sellers' business or an anomalous behavior. There can be several other reasons like - supporting someone's business, who is personally very close, or, buying similar product from some expert and again selling to someone else at a higher price. In 36.84\% of the cases, we find that buyer's total sales is more than the seller. In 7.81\% cases they offer gigs in the same category. These cases may be those where a high income seller supports other less popular sellers by promoting them or, by just providing good reviews for their gigs. If we consider the level wise distribution, then higher level sellers buy gigs from lower level sellers in 24.25\% cases, whereas, lower level sellers buy gigs from higher level sellers in 15.09\% cases. The remaining 60.66\% cases are purchases within the same level.
\vspace{-3mm}
\subsubsection{Reciprocity and revenue difference between buyer and seller}
Here, we test the hypothesis whether the average revenue difference between a seller and a buyer in a community decreases as the reciprocity increases. We calculate reciprocity for each of the communities of the network and then calculate the average difference of revenue for each of the communities. So, if a community $c_i$ of the network contains $\{(s_1,b_1),(s_2,b_2),\cdots , (s_N,b_N)\}$ with $s_j$ being the seller and $b_j$ being the buyer, we define average revenue difference of $c_i$ as
\begin{center}
$R_{i}$ = $\sum_{j=1}^{N}(|revenue(s_j)-revenue(b_j)|)/N$
\end{center}
Our test cases then are - 
%\todo{How is the reciprocity measured? Whether a $b_i$ has purchased something from $s_i$ and vice versa. If so, mention this clearly or state that this information is take from the First level seller-seller network.} 
\\
$H_0$: Avg. difference of revenue in a community decreases as reciprocity increases.\\
$H_1$: Avg. difference of revenue does not decrease with increasing reciprocity.
\\
We apply linear regression to check the dependence between these two factors. The correlation between the two factors comes out to be $-0.1538$ with the $p$-value of $0.0921$. Hence, with $90\%$ confidence we can reject our alternative hypothesis (i.e., $H_1$).\\
Hence, our study shows that if reciprocity increases i.e., if more people indulge into a two way traffic, then the average revenue difference decreases i.e., their sales values become almost identical. This means that when two sellers in the market almost perform similarly, they tend to support each other. This type of mutual uplifting can promote both their businesses.
\subsubsection{Effect of location on the volume of sales}
Intuitively one would expect that in a global marketplace, if two people belong to same geographical location, then their interaction would be much higher on average than with other people from different locations. To check this hypothesis for Fiverr, we calculate the average volume of sales happening in a particular community and check its dependence on the average geographical location similarity. For each community $c_i$ in the network, we define average location similarity as
\begin{center}
$L_i$ = $\sum_{j=1}^{N}(G_{s_j, b_j})/N$
\end{center}
where, $G_{s_j , b_j} = 1$ if both seller $s_j$ and buyer $b_j$ (both of them are from $c_i$) belong to the same geographical location and 0 otherwise. The correlation coefficient between average volume of sales and $L$ comes out to be $0.1871$ with a $p$-value of $0.0398$. Therefore, with a confidence of $97\%$ we can conclude that there is a positive correlation between our derived location similarity metric and average volume of sales. Thus, it is fair to assume that our claim, that users from the same location intend to purchase from one another and their volume of interaction is high, is reasonable.
\subsubsection{Effect of similar gigs on reciprocity}
Next, we calculate the average similarity between the gig names of the corresponding gigs provided by the seller and the buyer. Following this, we calculate the average gig similarity in a community. We apply linear regression to check the relationship between average gig similarity and reciprocity. Our results seems to strongly favor our claim since, as the average similarity between gigs increases, so is the propensity to reciprocate. The regression coefficient is $0.6447$ with a $p$-value of $0.00046$. Thus, if two sellers sell similar type of gigs then they are more inclined to buy from each other. This type of phenomenon is very rare in common marketplaces. If a buyer sells some product, then it is very unlikely that she will buy the same product from a different seller. It is even more unlikely that the second seller will again buy back some similar product from the first one.
\subsubsection{Relationship between seller level and reciprocity}
We observe that in $\sim60\%$ cases, seller-seller interactions happen between same level sellers. Therefore, the next question that arises is - if two same level seller interact, what is the probability that they reciprocate? \\
To answer this, we calculate average seller similarity between people in each of the communities. For each community $c_i$ we define avg. seller level similarity as
\begin{center}
$SR_i$ = $\sum_{j=1}^{N}(S_{s_j, b_j})/N$
\end{center}
where, $S_{s_j, b_j} = 1$ if $s_j$ and $b_j$ belong to same level (\textit{level 1} or, \textit{level 2} or, \textit{top rated}) and 0 otherwise. After applying statistical test of significance, we find that the correlation coefficient to be $0.2657$ with a $p$-value of $0.00322$. This result clearly shows that the reciprocity of a community increases with the average level similarity i.e., two people from same level have high propensity to reciprocate to each other.
\vspace{-2mm}
\subsection{Tier-II seller-seller network}
We form another network on the basis of number of buyers common between two sellers and denote this network as Tier-II network. We take the top sellers and draw an edge between a pair of sellers if they share any common buyer. We define the weight of the edge as -
\begin{center}
$weight(a,b) = |buyer_a \cap buyer_b|/|buyer_a \cup buyer_b|$
\end{center}
with $buyer_i$ being the set of buyers of the seller $i$.
Analyzing this network gives us insight about loyalty of buyers. For the subcategory \textit{Creative Logo Design}, the average degree of the 282 nodes in the network comes out to be 110.28 with only one strongly connected component. This is indicative of divided loyalty or no loyalty of buyers in Fiverr market. Moreover, the network follows small world properties with a diameter of 3. We also find that the network is highly connected with an average clustering coefficient of 0.34. These results also indicate the fact that Fiverr is a marketplace where the competition is less monopolistic, i.e., most of the sellers sell similar types of gigs in order to gain revenue and customers do not discriminate well among these gigs.
\vspace{-4mm}
\section{Fiverr under strategy lens}
\vspace{-4mm}
In this section, we evaluate Fiverr as a marketplace from strategy perspective focusing on the inter-entity dependence. For this study, we only choose those sellers whose sales are more than 500 per gig ($\sim2.45\%$ of the whole seller community).
\vspace{-2mm}
\paragraph*{Dependence between no. of gigs sold and revenue}
\vspace{-2mm}
Intuitively, a seller selling more gigs seems more likely to generate more revenue. We perform statistical test of significance on data of top sellers. We apply linear regression with number of gigs as the independent variable and obtain a regression coefficient of $0.2479$ with a $p$-value of $2.2 \times 10^{-16}$. This shows that the relationship between the two is weak for Fiverr.
\vspace{-2mm}
% \subsubsection*{Dependence between subcategory span and sales}
% We checked the correlation between number of subcategories in which a given seller is selling and the average sales generated out of them. The correlation between these two factors came out- to be $-0.04496$ with a $p$-value $0.0229$. Hence, we observe that there is not much linear dependence between these two. Hence, we conclude that most of the top sellers prefer to offer multiple gigs but in the same subcategory i.e, without diversifying their portfolio much in terms of the number of categories, they try to offer mostly similar types of services. We will analyze the similarity between gigs and their effect on other market characteristics to a greater extent in a later subsection.
% \vspace{-1mm}

\begin{figure}[h]
    \centering
\begin{minipage}[b]{0.42\columnwidth}
\includegraphics[width=0.95\linewidth, height=30mm, angle=0]{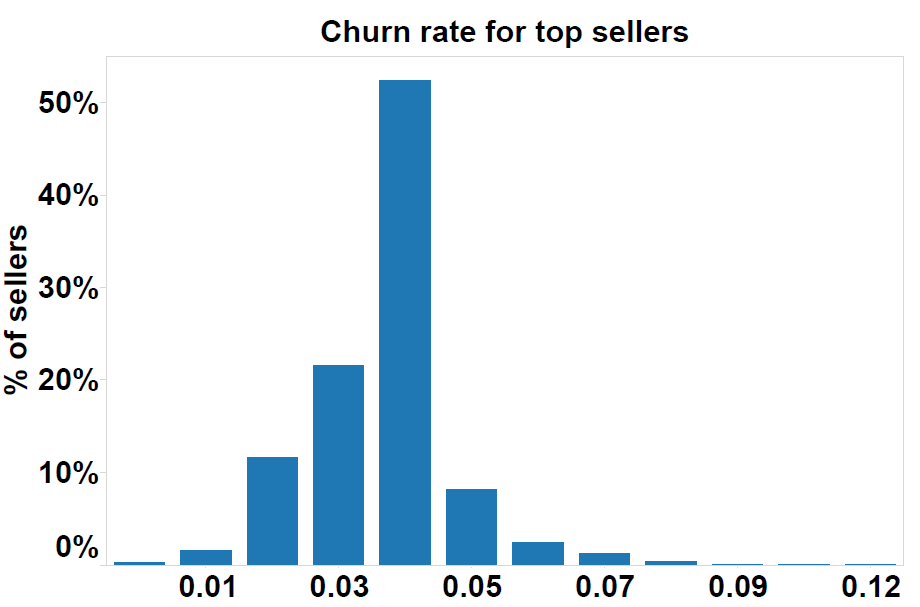}
\caption{\label{figchurn} Churn rate of top 2.5\% of sellers.}
\end{minipage}
\begin{minipage}[b]{0.42\columnwidth}
\includegraphics[width=0.95\linewidth, height=30mm, angle=0]{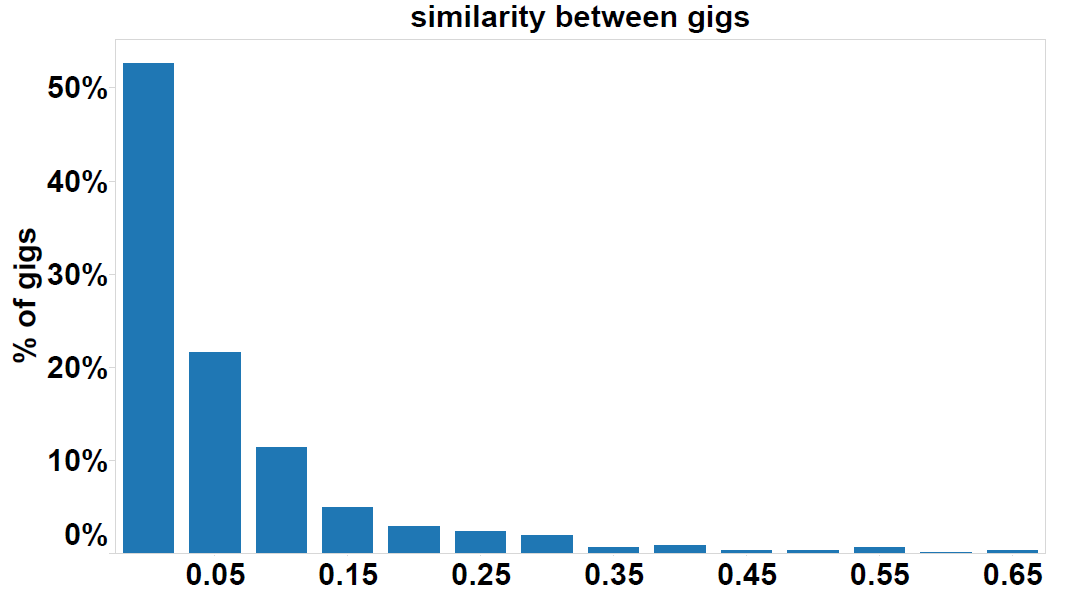}
\caption{\label{figsim} Inter-gig similarity of seller.}
\end{minipage}
\vspace{-4mm}
\end{figure}

\paragraph*{Customer churn rate and sales}
\vspace{-2mm}
We define customer churn rate for a seller for a particular gig as
\begin{center}
$churn$ $rate$ = $\sum_{t=1}^{N}{\frac{|\overline{customer_{t}} \cap  customer_{t-1}|}{|customer_{t-1}|}}$
\end{center}
Here, $customer_t$ is the set of the customers of the seller at time period $t$ and $N$ is total number of time periods the gig has been selling for in the last nine months.\\
We calculate the average churn rate for all gigs offered by a seller (fig~\ref{figchurn}). 
We believe that the lesser the customer churn rate, the better the customer handling and ability to satisfy a customer, and more the sales. This is of course not true for products with long life-cycles. For example, a customer who bought a mobile phone from a seller in Amazon is less likely to return for another phone purchase. In low value service driven marketplaces, customers returning can be a very frequent phenomenon. We studied the relationship between churning rate and total sales of Fiverr. 
Churn rate is somewhat affecting the total sales, and the two are positively related (correlation coefficient of 0.12). One reason could be that most of the customers of Fiverr are just one time buyers. Average churn rate is higher for many of the sellers.
% \begin{table}[h]
% \centering
% \label{tablecoeff}
% \begin{tabular}{lllll}
% \textbf{Corr. coeff.} & \textbf{Error} & \textbf{t-value} & \textbf{p-value} \\
% $0.1194$ & $0.04305$ & $2.774$ & $0.00573$\\
% \end{tabular}
% \end{table}
% \begin{figure}[h]
% \begin{center}
% \includegraphics*[scale=0.25, angle=0]{churn_rate_of_Top_sellers.PNG}
% \caption{\label{fig6666} Histogram showing the churn rate of top  2.5\% of sellers.}
% \end{center}
% \end{figure}
\vspace{-2mm}
\paragraph*{Dependence between similar gigs and sales}
\vspace{-3mm}
We observe that many top sellers sell more than one gig and in many cases they provide their services in the same subcategory. So, the next question that arises is - how similar are their gigs or do they provide different type of services? We create tf-idf matrices for 1-gram, 2-gram and 3-gram similarity of the gig names of the gigs the seller offers. As most of the gigs start with the phrase ``I will do'' or ``I will make'', we remove these phrases and other stop words from the gig names. After that, we take the average of the three tf-idf matrices to get one matrix $M$. Standardized rows of $MM^{T}$ gives us the average similarity between one gig and other gigs. Hence, average of all the off diagonal elements of the matrix $MM^{T}$ gives the average gig similarity of the particular seller. We then study the distribution of the similarity value (fig~\ref{figsim}) and check the significance of similarity value over number of sales made by a seller. The result we obtain from the statistical test of significance is a coefficient of $-0.216$ with a low $p$-value of $0.0073$. This results clearly show that the similarity value is negatively related to total sales i.e., lower the similarity value, higher are the sales. The more diverse type of products a seller sells, the more sales he/she can make.
\vspace{-3mm}
\paragraph*{Similarity with other gigs and sales}
In the previous subsection, we define the similarity between gigs of each seller. Here we define a similar concept, between a gig and other gigs. Therefore, we focus on the average similarity between a single gig with other gigs in the same category offered by other sellers. In a competitive market, we usually believe that the products which are somewhat different from the other products stand out and make good sales. For example, most of the niche products like Apple iPhones or automobile companies like BMW provide a unique experience to their customers which creates a strong brand awareness among the customers. In case of Fiverr, we wish to examine whether a unique gig that has very low similarity with other products in the category is having more sales, or the other way round. The correlation coefficient between average similarity with other gigs and sales is $0.1336$ with a very low $p$-value of $0.000027$. Very low $t$-statistics and a positive correlation clearly shows that there is a significant dependence between these two. Moreover, in Fiverr, higher similarity with others means more prospects of being successful. This tells us that most of the products offered in Fiverr are homogeneous in nature and the number of alternative products is also very high.\footnote{\url{https://en.wikipedia.org/wiki/Perfect_competition}}
\vspace{-3mm}
\paragraph*{Assessing competition in subcategories}
We define a similar concept of similarity between all gigs in each of the subcategories. If the average inter-gig similarity is high for a particular subcategory, that means most of the sellers tend to sell similar types of gigs in that category. This indicates that the prospect of being successful in that category is high, if one seller goes with the trend. This implies that very high competition exists in that subcategory. We observe whether this measure of competition affects the sales of a category. The topmost competitive subcategory came out to be \textit{Video Post Production Editing} with average sales per gig of 1,141. We standardize the total number of sales by the number of gigs in that subcategory to get the average sales per gig in that subcategory. After applying linear regression analysis on the two standardized variables, average sales per gig and average similarity between gigs, we get a high correlation coefficient of $0.4395$ with a very low $p$-value of $3\times 10^{-6}$. This result clearly shows that as the competition or the similarity between gigs in a subcategory increases, the average sales also increases. This type of phenomenon is very common in an ideal scenario of perfectly competitive market. Also, large number of buyers and sellers, accurate and necessary information about all the products, very flexible and barrier free market structure has made Fiverr a very good example of perfectly competitive marketplace. The average gig price which is very close to the actual market price of \$5 shows that every entity is a ``price taker'' in the market. %In the above subsection we have also show that monopolistic and unique sellers are very less in number in Fiverr, which again supports our claim.
\vspace{-3mm}
\paragraph*{Evaluation of growth prospective of seller in market}
Evaluation of performance of sellers is vital for any crowdsourcing platform. We use the classical method of Boston Matrix (BCG matrix\footnote{\url{https://en.wikipedia.org/wiki/Growth-share_matrix}}) to evaluate the performance of the different top 2.5\% sellers. The BCG matrix allows organizations to identify product positioning in the market and thus they could allocate resources accordingly. We also use the same concepts in order to classify the sellers in the similar lines. The different groups in BCG matrix are \\  
1. Stars : The best performing products or, sellers which generate most of the revenue and have positive growth rate and use most of the resources of the organization.\\ 
2. Question Mark : The products or, sellers having very high growth prospective but currently low market share. They are the potential stars.\\
3. Cash Cow : The market leaders with low growth rate. People invest in them to maintain their status in market.\\
4. Dogs : The products or, sellers with low market share as well as low growth prospects.\\
\begin{figure}[h]
\begin{center}
\includegraphics*[scale=0.16, angle=0]{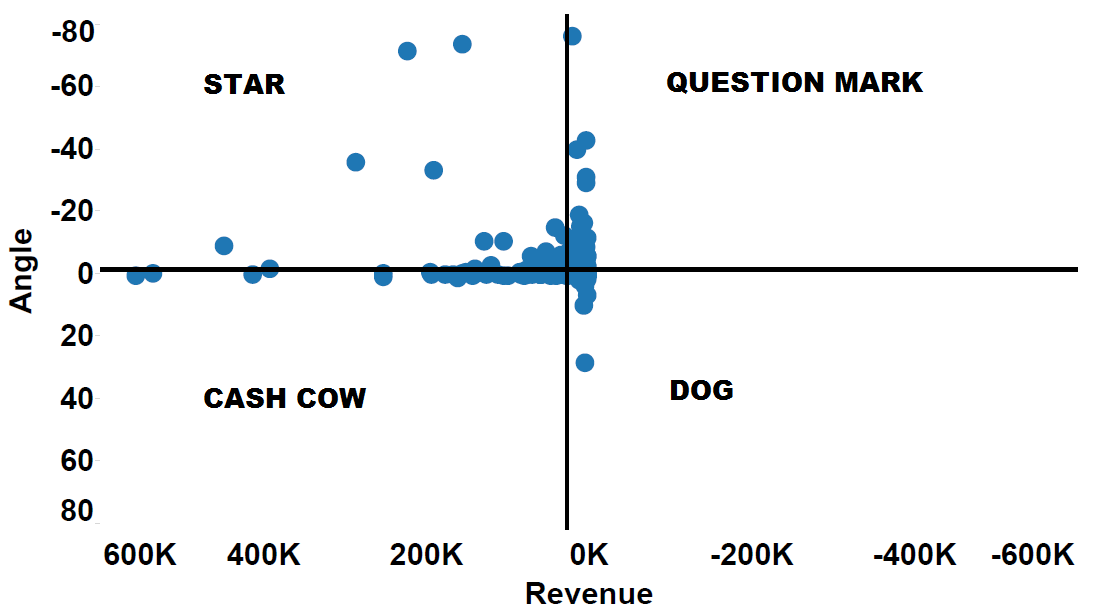}
\vspace{-3mm}
\caption{\label{figbcg} Boston Matrix showing sellers in different quadrants of matrix.}
\vspace{-4mm}
\end{center}
\end{figure}
\vspace{-1mm}
Based on the matrix (fig~\ref{figbcg}) we also divide top sellers in four groups - Stars, Cash cows, Question marks and Dogs based on their market share (i.e., proportion of revenue generated) and their growth rate measured as the slope of gain of revenue in consecutive months. Therefore, if a seller had revenue $\{r_1, r_2, \cdots ,r_N\}$ respectively in the time period $\{t_1, t_2, \cdots ,t_N\}$, then the growth of revenue for the time period $\{t_1, t_2, \cdots ,t_{N-1}\}$ would be $\{r_1 - r_2, r_2 - r_3, \cdots ,r_{N-1} - r_N\}$. We normalized the growth by the number of days and applied linear regression to get the slope of change of revenue growth. We then use the regression coefficient as the growth rate for each of the sellers. More negative the slope is, more is the growth rate of the seller.
\begin{table}[h]
\centering
\label{tabbcg}
\vspace{-3mm}
\resizebox{7cm}{!}{
\begin{tabular}{ll}
\textbf{Quadrant of BCG matrix} & \textbf{percentage of top sellers} \\
Stars  & 4.3\%\\
Cash cow  & 16.85\%\\
Question Mark  & 10.11\%\\
Dogs  & 68.74\%\\ 
\end{tabular}}
\vspace{-4mm}
\end{table}
The below table shows that there is very low percentage (4.3\%) of stars which is a common observation in any market place. Moreover, we see that the top performers in the \textit{Stars} have very high growth rate (slope is close to $-80^{\circ}$). Similarly, 16.85\% of the sellers have a growth rate close to zero but with high revenue. We further observe that 37.9\% of these people are from \textit{top rated} level. We believe that these sellers may have reached their saturation point and possibly will be taken over by \textit{level 2} in the next few months. Similarly, among the \textit{Questions Marks} 61.5\% are \textit{level 2} sellers who have a very high growth prospect with relatively low revenue in the current period. Providing proper incentive to these sellers can help them in becoming stars. Thus, Boston matrix provides the insights about incentivizing the right set of sellers in order to boost their businesses.
\vspace{-4mm}
\section{Discussions and conclusions}
\vspace{-4mm}
In this paper, we study the characteristic properties of Fiverr marketplace from various points of view - economic, sociological and strategy and make the following key observations.

The Fiverr marketplace is a unique in that the buyers who purchase gigs from sellers can convert themselves to sellers at any point in time.
Most sellers are proficient in certain type of services which indicates a good level of professionalism in the marketplace. Hence most of the top sellers prefer selling small number of gigs in a few subcategories rather than offering diverse category of gigs. From the viewpoint of the average gain of top sellers, the trends decrease over time. Globally, the percentage gain of customers for top sellers is around 40\%. However, some subcategories also show different trends. The gains sometimes reach as high as 75\%. The Fiverr marketplace is dominated by one-time buyers which consists of 45\% of the total buyers. An interesting and unique characteristic in a supply-driven marketplace is the existence of loyal customers. It is observed that many buyers buy the same product repeatedly from a few categories. This also results in the amount of purchases not always being directly proportionate to the volume of purchases. We observe that sellers are more appeasing in their interactions and try to woo their buyers into buying their gigs. Serious categories have comparatively more reviews with near zero and negative polarities as compared to the trivial categories. 

On analyzing the Tier-I seller-seller interaction network, we observe that there are many small tightly-knit seller communities existing in the network; however the overall network structure indicate loose connectivity with density of 0.001 among 49.13\% sellers. The very low clustering coefficient in the network shows that there are very few or no transitive links in the network which is quite uncommon for general markets. There is also low reciprocity in the network.

We also observe that some sellers buy from other sellers within the same subcategory. The possible reasons for this behavior include promoting which is supplemented by the buyer having more sales than the seller and providing positive reviews to the seller. Most of the top sellers prefer to offer multiple gigs but in the same subcategory. An analysis of the gigs of the top sellers indicates that the more diverse type of products a seller sells, the more sales he/she can make. In terms of gig similarity on Fiverr, higher similarity with others means more prospects of being successful. This tells us that the products offered in Fiverr are homogeneous in nature and the number of alternative products is also very high. Similarly, as the competition or the similarity between gigs in a subcategory increases, the average sales also increases. This type of phenomenon is very common in an ideal scenario of perfectly competitive market. Boston Matrix analysis shows that top performers in the `Stars' group have very high growth. At the same time, about ${\frac{1}{6}}^{th}$ of all the sellers have almost no growth but high revenue. Similarly, most of the sellers in the `question mark' bracket can potentially become stars if provided with the right incentives.

\vspace{-4mm}
\section{References}
\vspace{-12.5mm}
\bibliographystyle{ieeetr}
\bibliography{ref}
% that's all folks
\end{document}